**Relativistic trajectories**

A. LOINGER

Dipartimento di Fisica, Università di Milano

Via Celoria, 16 – 20133 Milano, Italy

**Summary.** – A quite evident and radical argument proves that no motion of masses can generate gravitational waves.

It is indisputable that if no motion of masses generates gravitational waves, the gravitational waves do not exist. Several physicists have proved irrefutably – and with different approaches – that no gravitational motion of pointlike masses gives origin to the emission of gravitational waves [1]. Unfortunately, the results of these authors have appeared unpalatable to the overwhelming majority of the astrophysicists, for unclear – or perhaps too clear – reasons.

I shall exhibit now a very easy, and yet perfectly logical, proof of the absence of emission of gravitational radiation in *any whatever* motion (gravitational or even non-gravitational) of a point mass.

Let us suppose, on the contrary, that at a given time $t$ of its motion a given point mass $M$ begins to send forth a gravity wave, and let us assume to know the *kinematic characteristics* of the motion between $t$ and $t + |dt|$. It is indubitable that we can reproduce these *same* characteristics for a gravitational motion of the mass $M$ in a suitable, fixed, "external" gravity field, within a suitably chosen time interval equal to $|dt|$. But in this case our mass moves along a **geodesic** line, and therefore it **cannot** emit any gravitational radiation. *Q.e.d.*

We see, in particular, how unfounded are those arguments which pretend to evaluate the gravitational power emitted in the motions of the heavenly bodies − for instance in the revolution motion of a planet or a binary star − by utilizing a well-known perturbative formula (indeed, of a very dubious conceptual validity). Thus, e.g., for the motion of Jupiter around the Sun one obtains a power of the order of magnitude of 450 watt − a paltry, and yet an illusive power, because the real power is exactly zero.

Conclusion: since no "device" exists for the production of a gravity wave (the above restriction to the motions of *pointlike* masses is clearly inessential), all the formal solutions of Einstein field equations which have an undulatory character [2] cannot describe *physical* waves.

(Of course, the astrophysical community should reconsider the research projects concerning the detection of the gravitational waves).

"An die Menge
Was für ein Dünkel! Du wagst, was wir alle loben, zu schelten?
Ja, weil ihr alle, vereint, auch noch kein Einziger seid".
                                            F.v. Schiller